\documentclass[paper]{JHEP3}
\usepackage{amssymb,enumerate}
\usepackage{amsmath}
\usepackage{bbm}
\usepackage{url}
\usepackage{cite}
\usepackage{graphics}
\usepackage{xspace}
\usepackage{epsfig}
\usepackage{subfigure}

\newcommand\POWHEG{{\tt POWHEG}}
\newcommand\POWHEGBOX{{\tt POWHEG~BOX}}
\newcommand\PYTHIA{{\tt PYTHIA}}
\newcommand\POWHEGpPYTHIA{{\tt POWHEG+PYTHIA}}
\newcommand\POWHEGpPS{{\tt POWHEG+PS}}
\newcommand\HERWIG{{\tt HERWIG}}
\newcommand\JIMMY{{\tt JIMMY}}

\def\({\left(} 
\def\){\right)} 

\def\beq{\begin{equation}}
\def\beqn{\begin{eqnarray}}
\def\eeq{\end{equation}}
\def\eeqn{\end{eqnarray}}

\def\mr{\mathrm}

\def\vbfww{VBF $W^+W^+jj$\;}
\def\wpp{W^+W^+}
\def\wmm{W^-W^-}
\def\evmv{e^+\nu_e\mu^+\nu_\mu}
\def\mc{\mathcal}

\title{NLO corrections to electroweak and QCD production of W$^+$W$^+$ plus two jets in the \POWHEGBOX} 
\vfill
\author{
Barbara J\"ager \\
Institut f\"ur Physik (THEP), Johannes-Gutenberg-Universit\"at, 55099 Mainz, Germany\\
E-mail: \email{jaegerba@uni-mainz.de}
}

\author{Giulia Zanderighi \\
Rudolf Peierls Centre for Theoretical Physics, 1 Keble Road, University of Oxford, UK\\
E-mail: \email{g.zanderighi1@physics.ox.ac.uk}

}

\keywords{POWHEG, NLO, QCD, SMC}

\abstract{We present the matching of the next-to-leading order QCD
  calculation for $\wpp jj$ production via vector-boson fusion in
  hadronic collisions to parton-shower Monte-Carlo programs according
  to the \POWHEG{} method. Our implementation complements existing
  code for QCD-induced $\wpp jj$ production in the \POWHEGBOX{}
  package, thereby providing a platform for the complete Standard
  Model production of $\wpp jj$ events via QCD and electroweak
  interactions.  The impact of parton-shower effects is discussed for
  various distributions and found to be small in most cases. However,
  few observables, that are relevant for analyses using a central jet
  veto, are modified significantly when they are interfaced to a
  parton shower program.}
\preprint{MZ-TH/11-20\\
  OUTP-11-45P}

\begin{document}

\section{Introduction}
With the start-up of the CERN Large Hadron Collider (LHC) the
production of multi-particle final states at high energies has become
feasible. The analysis and interpretation of these measurements
requires accurate predictions for processes of high multiplicity.
The past years have seen an exceptional progress in the development of
new techniques suitable for the calculation of next-to-leading order
(NLO) QCD corrections to multi-particle production processes at hadron
colliders (see \cite{Ellis:2011cr} for a pedagogical review on these
techniques).  These advances resulted in a large number of NLO
calculations of multi-particle production processes in the last two
years~\cite{Bredenstein:2010rs,Bevilacqua:2009zn,Berger:2009ep,Ellis:2009zw,KeithEllis:2009bu,Binoth:2009rv,Bevilacqua:2010ve,Berger:2010vm,Jager:2009xx,Bozzi:2007ur,Jager:2006cp,Jager:2006zc,Melia:2010bm,Dittmaier:2007wz,Denner:2010jp,Bevilacqua:2010qb,Berger:2010zx,Campanario:2011ud}.
While NLO-QCD predictions are essential to reduce the theoretical
uncertainties associated with the hard scattering process, a
description of the additional hadronic activity in realistic LHC
events relies on parton-shower generators such as
{\HERWIG}~\cite{Marchesini:1991ch,Corcella:2000bw} or
{\PYTHIA}~\cite{Sjostrand:2006za}.  
The perturbative accuracy of these programs is however limited,
formally, to leading logarithmic accuracy. 
Combining the benefits of an NLO-QCD calculation with those of a
parton-shower program yields the most realistic, yet accurate, results
for scattering processes at hadron colliders feasible to date.

Currently, two frameworks are available to combine NLO-QCD
calculations with parton shower programs: 
{\tt  MC@NLO}~\cite{Frixione:2002ik} and
\POWHEG~\cite{Nason:2004rx,Frixione:2007vw}. 
{\tt aMC@NLO} is a recent further development of the former approach
that aims at a fully automated event generation at NLO in QCD
\cite{Hirschi:2011pa}. A first phenomenological application of {\tt
  aMC@NLO}, has been the implementation of a (pseudo)-scalar Higgs in
association with a $t\bar t$ pair \cite{Frederix:2011zi}.
The \POWHEG{} method has been utilized by several groups, e.g., in the
context of \HERWIG++~\cite{Bahr:2008pv,D'Errico:2011sd}, or {\tt
  Sherpa}~\cite{Gleisberg:2008ta,Hoche:2010pf}. Recently, the method
has been used also for $t\bar t$ pair production in association with
one jet \cite{Kardos:2011qa} and in association with a Higgs boson~\cite{Garzelli:2011vp} in the HELAC framework.

As a tool for merging NLO-QCD calculations for arbitrary processes
with any parton shower program in the \POWHEG{} approach, the
so-called \POWHEGBOX~\cite{Alioli:2010xd} was developed one year
ago. It requires few process-specific building blocks: the Born and
real-emission squared amplitudes, the finite parts of the virtual
squared amplitude, color- and spin-correlated Born amplitudes, the
Born phase-space and the flavor structure of all contributing Born and
real subprocesses. All these elements are essentially readily
available once a NLO calculation for a given process has been
performed. Thus the implementation in the \POWHEGBOX{} of a new
process that is known at NLO requires in general only little effort.
All further ingredients for the merging procedure and the subtraction
of singularities are provided by the \POWHEGBOX. Using this method, a
large number of NLO-QCD calculations has been matched to parton-shower
programs recently and is currently in the public repository of the
\POWHEGBOX{}: heavy flavor pair production~\cite{Frixione:2007nw},
Higgs production via gluon fusion~\cite{Alioli:2008tz} and via vector
boson fusion (VBF)~\cite{Nason:2009ai}, vector
boson~\cite{Alioli:2008gx} and vector boson plus jet
production~\cite{Alioli:2010qp}, single top
production~\cite{Alioli:2009je,Re:2010bp}, jet pair
production~\cite{Alioli:2010xa}, $Wb\bar b$
production~\cite{Oleari:2011ey}, vector boson pair
production~\cite{Melia:2011tj}, 
and QCD-induced $\wpp jj$ production~\cite{Melia:2011gk}.

This last process, $\wpp jj$ production, exhibits several
interesting phenomenological features. Due to its 
exotic signature with two same-sign leptons, missing energy, and two
jets it comprises an important background to double parton
scattering~\cite{dps,Gaunt:2010pi}. Understanding this class of
processes is of paramount importance for a realistic description of
the collider environment in which hard scattering reactions take
place. Moreover, $\wpp jj$ production constitutes a background for new
physics scenarios involving same-sign leptons, such as R-parity
violating SUSY models~\cite{Dreiner:2006sv} or doubly charged Higgs
production~\cite{Maalampi:2002vx}.  

In the Standard Model (SM) there are two direct hard production
mechanisms that give rise to the same $\wpp jj$ final state:
the QCD production, which proceeds via gluon-mediated
(anti-)quark scattering processes with the two $W^+$ bosons being
radiated each off one of the fermion lines (this is the process
already included in the \POWHEGBOX{}), and electroweak (EW)
production, where a color neutral vector boson connects the two quark
lines and the $W^+$ bosons are emitted either from the quarks or from
the exchanged vector boson.

At leading order, due to color conservation, the QCD and EW production
modes do not interfere and can thus be discussed separately.
Because of the hierarchy between the QCD and the EW coupling, one
would expect the QCD-induced $\wpp jj$ production, ${\cal
  O}(\alpha_s^2 \alpha^4$), to fully dominate over the EW production
mode, ${\cal O}(\alpha^6)$.
Explicit calculations revealed, however, that the inclusive QCD
production cross section of $\wpp jj$ final states is only roughly
twice as large as the corresponding EW one \cite{dps}. This
means that an accurate description of the inclusive SM production of $\wpp
jj$ must take both production modes into account.
Furthermore, when typical VBF cuts are imposed on the final state, the
QCD-production cross-section drops drastically and the SM
cross-section is fully dominated by the EW production mode.

NLO-QCD predictions for EW $\wpp jj$ production have been available
for some time~\cite{Jager:2009xx}. More recently, QCD corrections have
been provided also for the QCD production mode~\cite{Melia:2010bm},
and elements of this latter calculation have already been implemented
in the \POWHEGBOX. To allow for a complete SM description of the $\wpp
jj$ final state at NLO-QCD accuracy with parton-shower effects, we
implement the EW production mode in the same framework too. The
calculation of Ref.~\cite{Jager:2009xx} includes non-resonant diagrams,
takes off-shell effects and spin-correlations of the leptonic $W$
decays fully into account.

For definiteness, results presented in this paper are based on the
decays of the $W$-bosons to $e^+\nu_e \mu^+\nu_\mu$. Neglecting
interference effects in the case of identical leptons, the
cross-section for the $W$-bosons to decay to any lepton ($e^+, \mu^+,
\tau^+$) can be obtained by multiplying our results by nine over two.

We also note that, while we discuss here $\wpp$ production, which is
dominant at the LHC, it is also trivial to obtain results for
$\wmm$. Indeed, using charge conjugation and parity, $pp \to \wmm jj$
is almost equivalent to $\bar p \bar p \to \wpp jj$, modulo reversing
the momentum directions for parity-odd distributions.

The outline of this paper is as follows: In Sec.~\ref{sec:tech} we
describe the technical details of the NLO-QCD calculation for \vbfww
production, its implementation in the \POWHEGBOX, and the checks we have
performed.  In Sec.~\ref{sec:pheno} we discuss phenomenological results for
several kinematic distributions with particular emphasis on the impact
parton-shower effects have on the NLO-QCD predictions. Cross sections
and distributions for the VBF production mode are compared to the same
observables of the QCD production mode. We conclude in
Sec.~\ref{sec:conc}.

\section{Technical details}
\label{sec:tech}

\subsection{Next-to-leading order QCD corrections to \vbfww production}
The partonic scattering amplitudes which are needed for the
calculation of cross sections and distributions at NLO-QCD accuracy
and their combination with parton-shower programs via the \POWHEGBOX{} are
extracted from previous work on \vbfww production in the {\tt vbfnlo}
framework~\cite{Arnold:2008rz}. Here, we summarize only the elements
of the calculation which are relevant for its implementation in the
\POWHEGBOX. For technical details on the NLO calculation, we refer the
interested reader to Ref.~\cite{Jager:2009xx}

At order $\alpha^6$, EW production mainly proceeds via the
scattering of two (anti-)quarks by $t$-channel exchange of a weak
gauge boson with subsequent emission of two $W^+$ bosons, which in
turn decay leptonically. Non-resonant diagrams where leptons are
produced via weak interactions in the $t$-channel also
contribute. Sample diagrams for both of these topologies are depicted
in Fig.~\ref{fig:vbf-lo} for the representative subprocess $uc\to\evmv
ds$.
\FIGURE[t]{
\includegraphics[angle=0,scale=0.6]{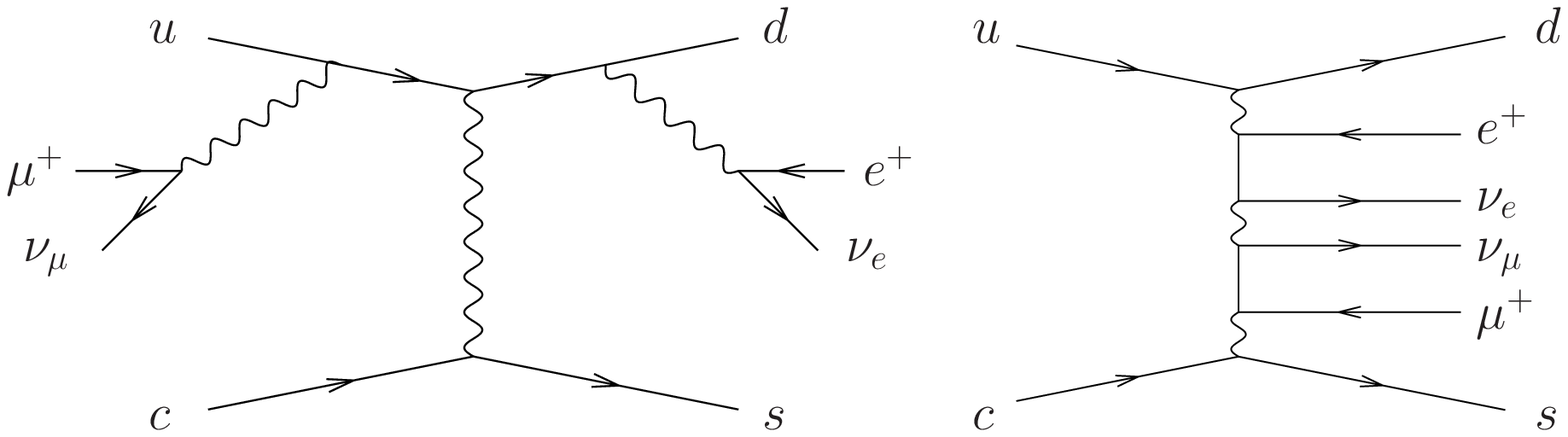}
\caption{Resonant (a) and non-resonant (b) sample diagrams for the
  partonic subprocess $u(1) c(2)\to e^+(3) \nu_e(4) \mu^+(5)\nu_\mu(6)
  d(7) s(8)$ at leading order.}
\label{fig:vbf-lo}
} 
In principle, also quark-antiquark annihilation processes with
weak-boson exchange in the $s$-channel and subsequent decay into a
pair of jets have to be considered. However, they are strongly
suppressed in the phase space regions where the VBF process can be
observed experimentally~\cite{Ciccolini:2007ec}, and thus
disregarded. For subprocesses with identical flavor combinations, such
as $uu\to\evmv dd$, in addition to the $t$-channel contributions
discussed above also $u$-channel diagrams and their interference with
the $t$-channel graphs arise. While we do take into account both $t$-
and $u$-channel contributions, we do not consider their interference
cross section, which is kinematically strongly
suppressed~\cite{Ciccolini:2007ec}.  For brevity, we will refer to
$pp\to e^+\nu_e\mu^+\nu_\mu jj$ as ``\vbfww'' production in the
following, even though the electroweak production process includes
non-resonant diagrams that do not stem from a $W^+$ decay (see
e.g. Fig.~\ref{fig:vbf-lo}b).

The NLO-QCD corrections to \vbfww production consist of two types of
contributions: one-loop corrections to the leading order (LO) diagrams and
real-emission contributions with an extra parton in the final state.
The real-emission amplitudes are obtained by attaching a gluon to the
tree-level diagrams discussed above. Crossing-related processes with a
gluon in the initial state also contribute. The virtual corrections
comprise the interference of one-loop diagrams with the tree-level
amplitude. Within our approximations, only selfenergy-, triangle-,
box-, and pentagon-corrections to either fermion line have to be
considered. Diagrams with a gluon being exchanged between upper and
lower quark lines vanish when interfered with the Born amplitude, due
to color conservation. The finite parts of the virtual corrections are
evaluated numerically by means of a Passarino-Veltman tensor
reduction. In order to avoid numerical instabilities in the evaluation
of the five-point tensor integrals, we resort to the Denner-Dittmaier
reduction scheme of Refs.~\cite{Denner:2002ii,Denner:2005nn}. The
numerical stability of our implementation is monitored by checking
Ward identities at every phase space point.
While in the NLO-QCD calculation of Ref.~\cite{Jager:2009xx} infrared
singularities arising in both real emission and virtual contributions
were handled by the dipole subtraction approach of Catani and
Seymour~\cite{Catani:1996vz}, the \POWHEGBOX{} internally takes care of
these divergences via the Frixione-Kunszt-Signer subtraction
approach~\cite{Frixione:1995ms}.

As in the NLO calculation in~\cite{Jager:2009xx} we neglect quark
masses, and entirely disregard contributions from external $b$ and $t$
quarks. We also assume a diagonal CKM matrix. 

\subsection{The \POWHEGBOX{} implementation}
We have implemented our NLO-QCD calculation for $pp\to \evmv jj$ via VBF  in the \POWHEGBOX{} by providing the following ingredients:
\begin{itemize}
\item
the list of all independent flavor structures of the Born process,
\item
the list of all independent flavor structures of the real emission process,
\item
the Born phase space,
\item
the Born squared amplitude,
\item 
the color-correlated and the spin-correlated Born squared amplitude,
\item
the real-emission squared amplitudes for all contributing sub-processes,
\item
the finite part of the virtual amplitudes interfered with the Born, 
\item
the color structure of the Born process in the limit of a large number
of colors. 
\end{itemize}
In order to comply with the ordering scheme required by the \POWHEGBOX, we
assign identifiers to the partons and leptons of each tree-level
subprocess in the following way:
\begin{itemize}
\item  1. incoming parton with positive rapidity,
\item  2. incoming parton with negative rapidity,
\item  3. -- 6. final state charged leptons and neutrinos,
\item  7. -- 8. final-state partons.  
\end{itemize}
For example, the particles involved in the representative subprocess
of Fig.~\ref{fig:vbf-lo} are ordered in the following way: $u(1)
c(2)\to e^+(3) \nu_e(4) \mu^+(5)\nu_\mu(6) d(7) s(8)$, where the
numbers in parentheses denote the identifiers. In the real-emission
configurations, the extra final-state parton is referred to as ninth
particle.

Spin-correlated Born amplitudes generally arise only if external
gluons emerge at tree-level, which is not the case in VBF processes.
Due to the simple color structure of the process, the only
non-vanishing color-correlated amplitudes $\mc{B}_{ij}$ for a pair of
partons $ij$ are just multiples of the Born amplitude $\mc{B}$ itself,
\beq
\mc{B}_{17} = \mc{B}_{28} = C_F \mc{B}\,,
\eeq
where $C_F=4/3$.  The assignment of color flow is straightforward and
unambiguous: it follows directly the propagation of the QCD partons.

In VBF processes, there is no interference between radiation off the
upper and lower fermion lines because of the color-singlet nature of
the weak-boson exchange in the $t$-channel. In order to pass this
extra information to the \POWHEGBOX{} while searching for singular
regions, a tag is assigned to each quark line, as explained in some
detail in Ref.~\cite{Nason:2009ai} for the related process of Higgs
production via VBF.  Particles that do not need to be tagged are
assigned a tag equal to zero, as illustrated in
Fig.~\ref{fig:vbf-tag}.
\FIGURE[t]{
\includegraphics[angle=0,scale=0.6]{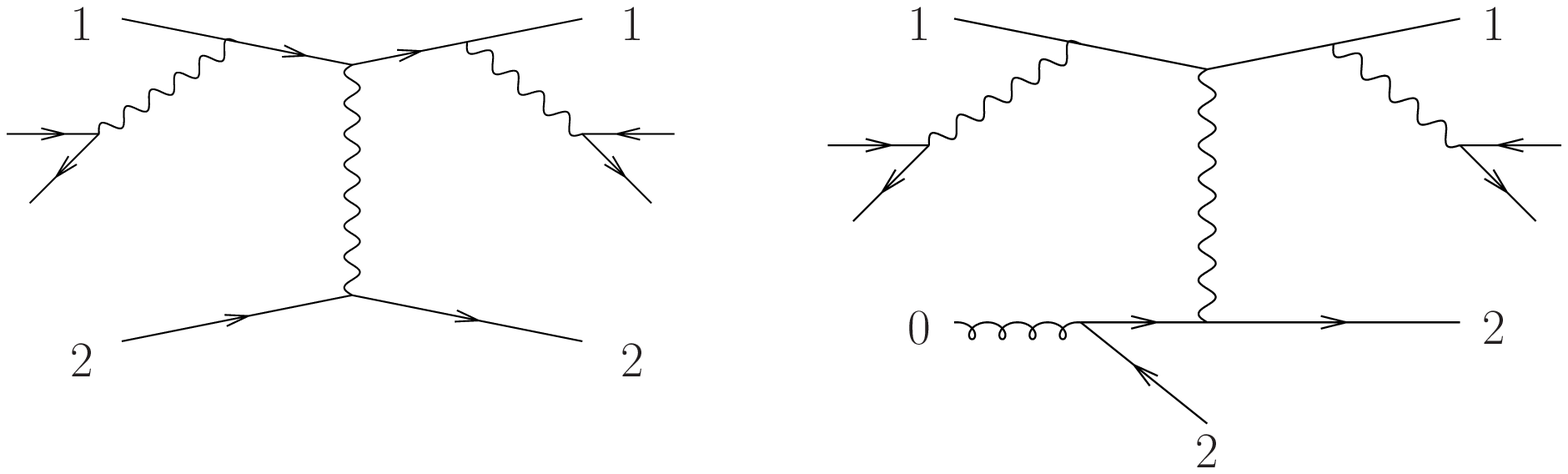}
\caption{Tag assignment for representative contributions to the Born
  process $uc\to e^+\nu_e\mu^+\nu_\mu ds$ and the gluon-initiated
  real-radiation process $ug\to e^+\nu_e\mu^+\nu_\mu ds\bar c$. Zero
  tags for the leptons are not explicitly indicated in the diagrams.}
\label{fig:vbf-tag}
}
The tags are taken into account by the \POWHEGBOX{} when singular regions
for the generation of radiation are identified, but have no further
consequences for the flavor arrangements.
The total LO cross section for $pp\to \wpp jj$ is finite. We therefore
apply a phase space generator that covers all phase space. Generation
cuts are not needed. In the code, $W$-bosons can be generated both on-shell or
off-shell with a Breit-Wigner distribution, according to the setting in
the {\tt powheg.input} file, see the {\tt VBF\_Wp\_Wp} manual for details.

We note that the implementation of the \vbfww process in the \POWHEGBOX{}
was straightforward. Following the procedure of
Ref.~\cite{Nason:2009ai} for the tagging of quark lines as described
above and exploiting the technical developments of
Ref.~\cite{Melia:2011gk} for the efficient treatment of a multi-leg
process, we were able to obtain a stable code without any further
systematic refinements.

\subsection{Validation and checks}
\label{sec:valid}
While the \POWHEGBOX{} offers a flexible environment for hadronic
scattering processes at NLO-QCD accuracy, the actual implementation of
the required building blocks requires care by the individual user and
has to be validated thoroughly.

A valuable check for the correct implementation of the tree-level and
real-emission contributions of individual subprocesses as well as
their flavor sum is provided by the \POWHEGBOX{} itself. It  
computes the soft and collinear limits of the real-emission cross
section internally, using only the user-supplemented Born
amplitudes. The \POWHEGBOX{} then checks automatically, whether these 
approach the full real-emission contributions in all soft and
collinear regions. The result of this comparison is written to a
user-readable file at the beginning of each production run.

The \POWHEGBOX{} does not only offer a framework for interfacing NLO-QCD
calculations to parton-shower programs, but can also be used in a
stand-alone mode for the calculation of cross sections and
distributions at LO and NLO-QCD accuracy. Using this option, we have
generated a multitude of distributions both at LO and NLO QCD, and
compared the results with those we obtained with the parton-level
Monte-Carlo program of Ref.~\cite{Jager:2009xx}. We found full
agreement for all observables.

\section{Phenomenological results}
\label{sec:pheno}
The implementation of \vbfww in the \POWHEGBOX{} has been made publicly
available. The interested reader can thus generate any required
distributions with all settings adapted to her needs. The default
setup of the code is identical to the QCD $\wpp jj$ production process
in the \POWHEGBOX, which has already been discussed in
Ref.~\cite{Melia:2011gk}. This should facilitate a comparison of the
two production modes is a common environment.
Here, we discuss only a few selected results for representative input
parameters and selection cuts. We restrict ourselves to the $\evmv jj$
final state with two charged leptons of different type and their
associated neutrinos.  The full cross section summed over electrons
and muons can be obtained thereof by multiplying with a factor of two
in the approximation that interference effects for same-type lepton
contributions are neglected.

We consider proton-proton collisions at a center-of-mass energy of
$\sqrt{s}=7$~TeV. For our phenomenological analysis, we use the NLO
set of the MSTW2008 parton distributions~\cite{Martin:2009iq},
corresponding to $\alpha_s(m_Z)=0.12018$.  As electroweak input
parameters we have chosen $m_Z=91.188$~GeV, $m_W=80.419$~GeV, and
$G_F=1.16639\times 10^{-5}~\mr{GeV}^{-2}$.  The other parameters,
$\alpha_\mr{QED}$ and $\sin^2\theta_W$ are computed thereof via LO
electroweak relations. The widths of the weak bosons are set to
$\Gamma_W = 2.099$~GeV and $\Gamma_Z = 2.51$~GeV. For the
reconstruction of jets from final-state partons, we use the {\tt
  FASTJET} implementation~\cite{Cacciari:2005hq} of the $k_T$
algorithm~\cite{Catani:1993hr,Ellis:1993tq}, with a resolution
parameter of $R=0.4$.

It is known that using as a scale just the mass of the weak boson
gives rise to larger, and shape dependent $K$-factors. In
Ref.~\cite{Jager:2009xx} local scales were used that can not be
implemented in a straightforward way in the \POWHEGBOX{}. This is
because the \POWHEGBOX{} currently intrinsically assumes that the
renormalization scales are the same at all vertices.
We therefore
set the factorization and renormalization scales here to
\beq
\mu_R = \mu_F =
\frac{p_{T,p1} +p_{T, p2} +E_{T, W_1} +E_{T, W_2}}{2}\,,
\qquad 
\mr{with}\quad
E_{T, W} = \sqrt{M_W^2+p_{T,W}^2}\,, 
\label{eq:scale}
\eeq
where the $p_{T, pi}$ are the transverse momenta of the two
final-state partons of the underlying Born configuration and each
$p_{T,W_i}$ represents the transverse momentum of a same-type
lepton-neutrino pair. In the public code, the user is free to
implement different scales and use any set of parton distributions
(PDFs) available from the LHAPDF library~\cite{Whalley:2005nh}.
Process-specific selection cuts can easily be modified in the analysis
routine.
  
As explained in some detail in Ref.~\cite{Frixione:2007vw}, in the
\POWHEG{} method the hardest emission is always generated first, with a
technique that yields positive-weighted events using the exact NLO-QCD
matrix elements. The \POWHEG{} output can thus be interfaced most easily
to a parton-shower Monte-Carlo program that is $p_T$ ordered, or
allows the implementation of a $p_T$ veto. \PYTHIA{}~6 does adopt
transverse-momentum ordering and thus can be interfaced in a
straightforward manner to \POWHEG. On the other hand, angular-ordered
parton-shower Monte Carlos, such as \HERWIG{}, may generate soft
radiation before generating the radiation of highest $p_T$. To account
for this type of radiation, a vetoed-truncated shower must be included
for a fully consistent matching with \POWHEG.
For our phenomenological analysis, we use
\PYTHIA{}~6.4.21~\cite{Sjostrand:2006za} to shower the events, include
hadronization corrections and underlying event with the Perugia 0
tune. In this study, QED radiation effects are not taken into account.
We have also showered events using {\tt
  HERWIG}~6.51~\cite{Corcella:2000bw}, supplemented by \JIMMY~4.31 for
the simulation of multiple parton interactions. While {\tt HERWIG},
utilizing an angular-order parton shower, does not provide the
truncated shower required for a fully consistent matching with \POWHEG{}
at double-logarithmic accuracy, the effect of neglecting the
associated contributions is expected to be small
\cite{Nason:2004rx}. Indeed, we found that for most observables the
results obtained by interfacing \POWHEG{} to \HERWIG{}+{\tt JIMMY} are very
similar to the \POWHEG{}+\PYTHIA{} predictions. In the following we
will thus focus on \PYTHIA{} results.

We require the presence of two jets with transverse momentum larger
than $p_{T,j} = 20$~GeV.  The two jets of highest transverse momentum
are referred to as ``tagging jets''.  We note, however, that the
cross-section is finite even without any cut on the jets, and that
since typical transverse momenta in the event are much larger than
$20$ GeV (see Fig.~\ref{fig:pttag}) the full cross-section without any
cut is very close to the numbers quoted in the following.  For the
scale choice in Eq.~\eqref{eq:scale}, we obtain an NLO
cross-section\footnote{Here and in the following, our errors are only
  statistical and do not include theoretical uncertainties such as
  scale or PDF dependencies of the results.} of $2.12(2)$~fb for the
QCD and $1.097(6)$~fb for the electroweak production mode. For our
set-up ($pp$ collisions at $7$ TeV) the QCD cross-section is thus
approximately twice as large as the electroweak one.  Therefore, as
observed already in Ref.~\cite{Gaunt:2010pi}, despite the hierarchy
between the strong and the electroweak coupling constant, the QCD
production cross-section is only moderately larger than the
electroweak one.

Furthermore, selection cuts that exploit characteristic differences
between the two production modes can serve as a powerful tool for the
suppression of QCD-induced $\wpp jj$ production, while they reduce the
electroweak production cross section only marginally.  Imposing such
cuts is particularly important in the context of searches for new
physics which manifests itself in the weak-boson scattering mode. In
order to identify such signatures, an efficient suppression of the QCD
background is mandatory.

Figure~\ref{fig:jets-nocuts}
\FIGURE[t]{
\includegraphics[angle=0,scale=0.5]{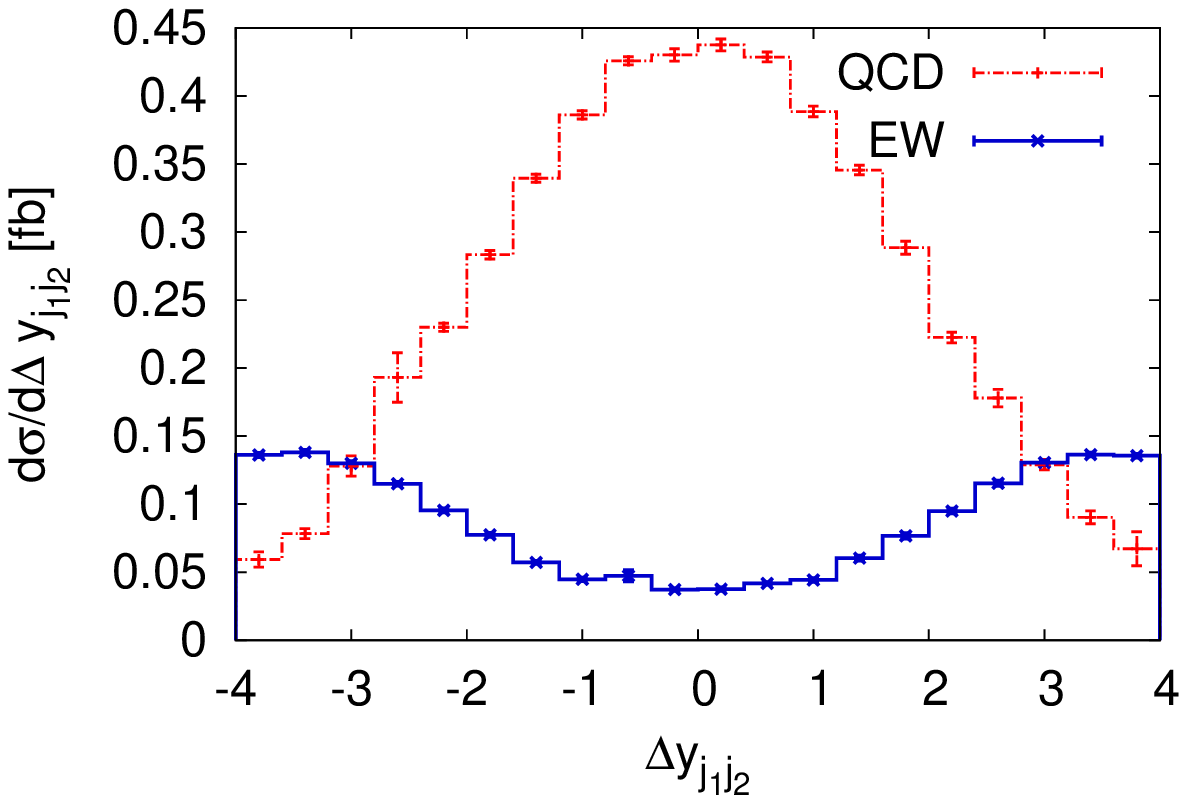}
\includegraphics[angle=0,scale=0.5]{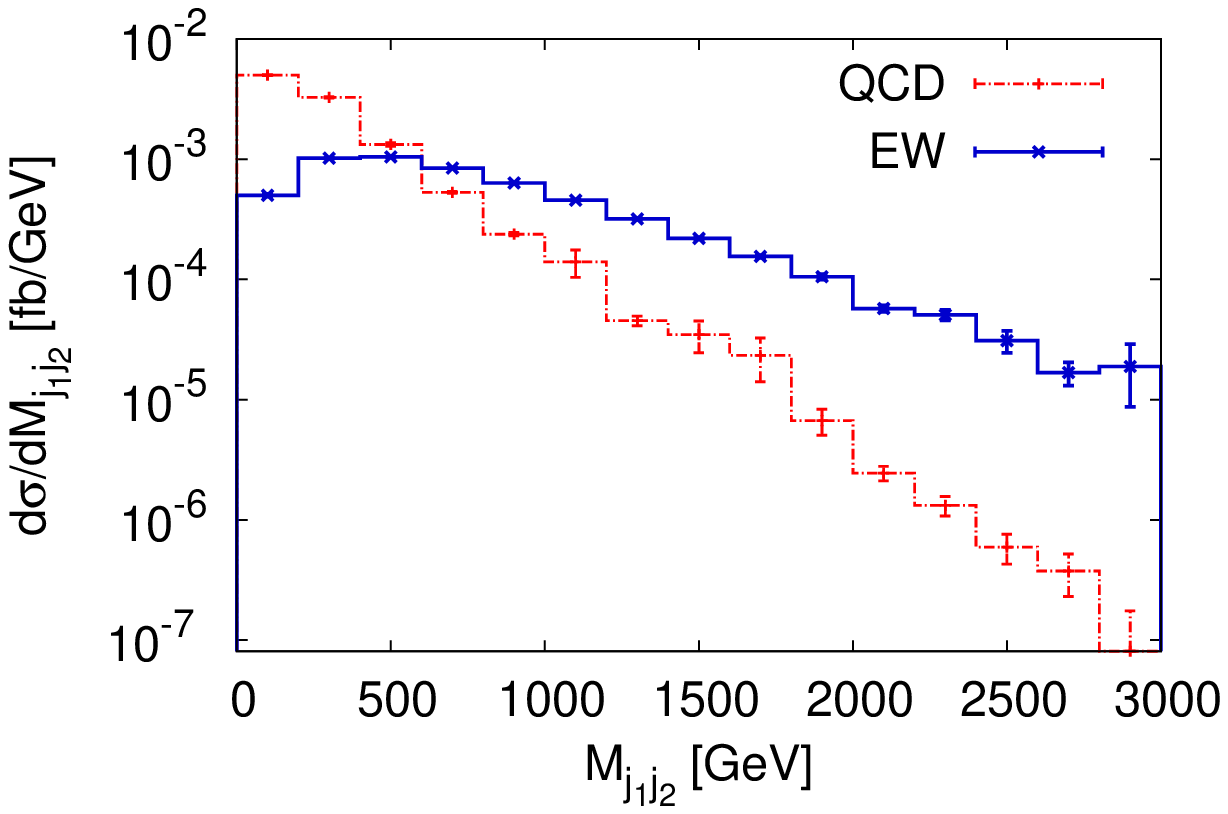}
\caption{Rapidity separation (left panel) and invariant mass (right
  panel) distributions of the two jets of highest transverse momentum
  at NLO-QCD accuracy for the QCD (dashed red lines) and EW (solid blue lines)
  production modes to $\evmv jj$ production at the LHC with $\sqrt{s}
  = 7$~TeV.  }
\label{fig:jets-nocuts}
} 
illustrates the most striking differences between the QCD and
electroweak production modes. Jets produced via QCD interactions tend
to be close in rapidity. Consequently, the distribution for the
rapidity separation of the two tagging jets, $\Delta
y_{j_1j_2}=y_{j_1}-y_{j_2}$, peaks at zero for the QCD $\wpp jj$
production mode.  On the other hand, \vbfww production gives rise to
two hard jets in the forward and backward regions of the detector,
which are well separated in rapidity. Requiring a large rapidity
separation of the two leading jets is thus an efficient tool for
enhancing VBF with respect to QCD contributions. Similar differences
can be observed for the invariant mass distribution,
$d\sigma/dM_{j_1j_2}$, of the two leading jets. Jet pairs of small
invariant mass are characteristic for the QCD production mode, while a
rather hard invariant mass distribution results from \vbfww
production. Imposing a large invariant-mass cut on the two leading
jets thus further reduces the relative impact of QCD contributions on
the $\wpp jj$ final state.

In the following we focus on a set of cuts which are specifically
designed to enhance electroweak with respect to QCD contributions,
making use of the aforementioned features of the two distinct
production modes. Since these selection criteria are particularly
useful in the phase space region where weak boson scattering is
searched for, they are usually referred to as ``VBF cuts''.
Throughout our phenomenological analysis, we consider jets as
identified only if their transverse momenta are larger than
\beq
\label{eq:ptj-cut}
p_{T,j}\ge 20~\mr{GeV}\,,
\eeq
in the rapidity-region accessible to the detector,
\beq
\label{eq:yj-cut}
y_j\le 4.5\,.
\eeq
Events with less than two jets fulfilling the criteria of
Eqs.~(\ref{eq:ptj-cut})~and~(\ref{eq:yj-cut}) are disregarded. 
The two tagging jets are required to be well-separated in rapidity,
\beq
|\Delta y_{j_1j_2}| = |y_{j_1}-y_{j_2}|>4\,,
\eeq
in opposite hemispheres of the detector,
\beq
y_{j_1}\times y_{j_2}<0\,,
\eeq
with large invariant mass,
\beq
M_{j_1j_2}>600~\mr{GeV}\,.
\eeq
On the two hard charged leptons, we impose the transverse-momentum cut
\beq
p_{T,\ell}\ge 20~\mr{GeV}\,,
\eeq
and demand that they be located in the central-rapidity region,
\beq
|y_\ell|\le 2.5\,.
\eeq
We furthermore require that they be well-separated from the tagging
jets and from each other in the rapidity-azimuthal angle plane,
\beq
\Delta R_{j\ell}\ge 0.4\,,\qquad
\Delta R_{\ell\ell}\ge 0.1\,.
\eeq
In addition, the charged leptons are supposed to fall between the two
tagging jets in rapidity,
\beq
\min\{y_{j1},y_{j2}\}<y_\ell<\max\{y_{j1},y_{j2}\}\,.
\eeq

With the above settings, we obtain an NLO-QCD cross section for the
VBF production mode of $\sigma^\mr{cuts} = 0.201(3)$~fb.
Therefore cuts reduced the electroweak cross-section by a factor 5. In
contrast, the NLO cross-section for the QCD production mode amounts now
to only $0.0074(7)$~fb, i.e.\ the cuts effectively reduced this
cross-section by a factor of almost 300. We conclude that once VBF
cuts are imposed, the QCD production mechanism becomes insignificant,
and we neglect it in the following.
After merging the VBF NLO matrix elements with {\tt PYTHIA}, the cross
section within the same cuts amounts to $0.1808(6)$~fb.  Clearly,
due to the extra hadronic activity generated by the shower, fewer
events pass the selection criteria that we impose. This change in the cross section should be kept in mind when very precise theoretical predictions are required, as is the case, for instance, in the context of coupling measurements in VBF processes at the LHC~\cite{Duhrssen:2004cv}. 

Even though the cross sections differ by around 10\%, the shape of the
transverse momentum distributions of the two tagging jets, displayed
in Fig.~\ref{fig:pttag},
\FIGURE[t]{
\includegraphics[angle=0,scale=0.5]{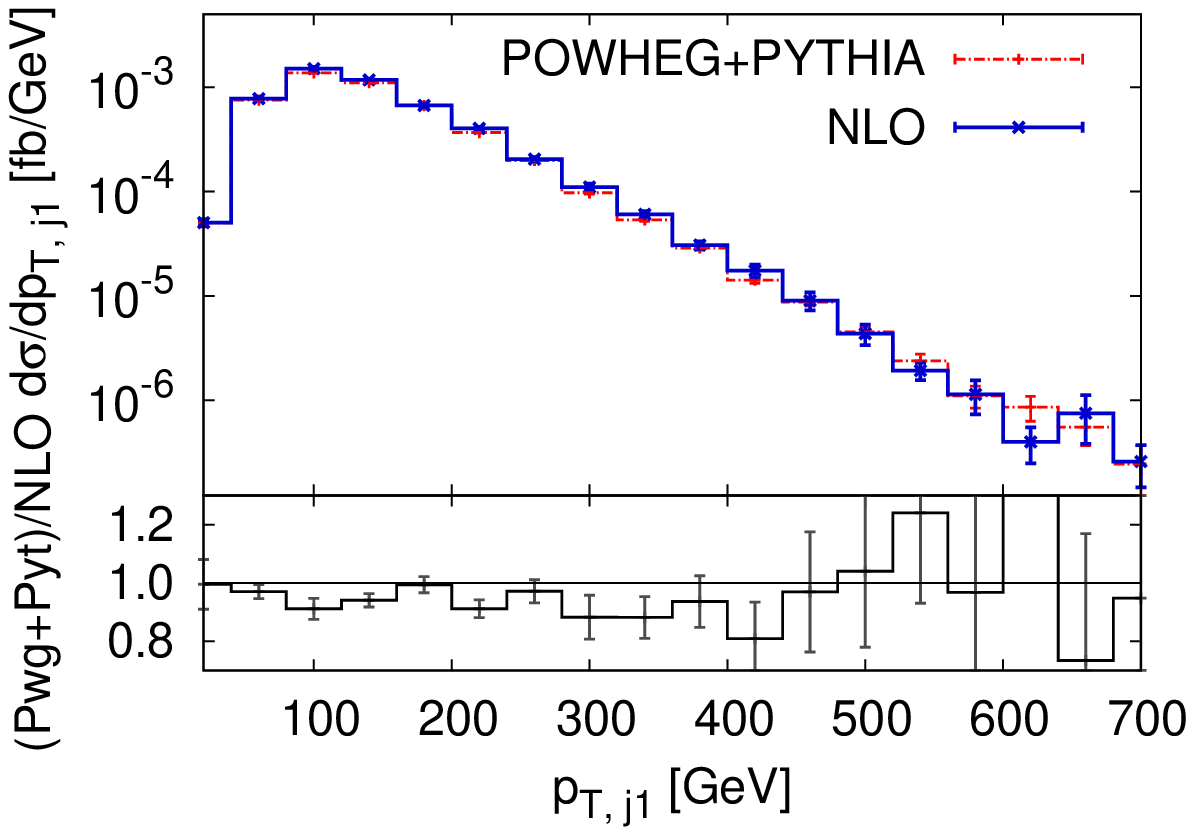}
\hspace*{1cm}
\includegraphics[angle=0,scale=0.5]{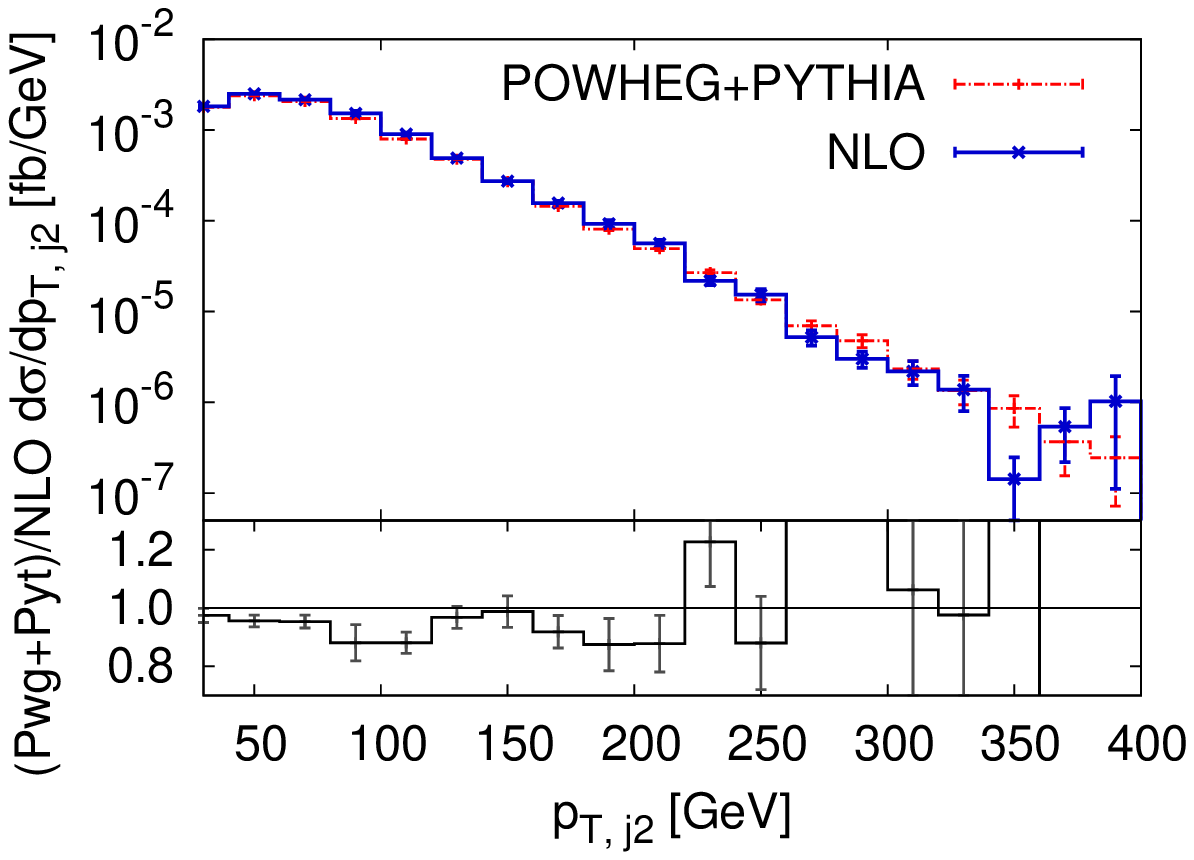}
\caption{Transverse momentum distributions of the two tagging jets at
  NLO-QCD (solid blue lines) and with {\tt POWHEG+PYTHIA} (dashed red
  lines) for $\evmv jj$ production at the LHC with $\sqrt{s} = 7$~TeV. 
  The lower panels show the respective ratios of the {\tt
    POWHEG+PYTHIA} to the NLO results.}
\label{fig:pttag}
} 
are hardly affected by parton shower effects. 
The production of tagging jets of high rapidity becomes somewhat less
likely in the {\tt POWHEG+PYTHIA} case, however, as illustrated by the rapidity
distributions of the two tagging jets in Fig.~\ref{fig:ytag}.
\FIGURE[t]{
\includegraphics[angle=0,scale=0.5]{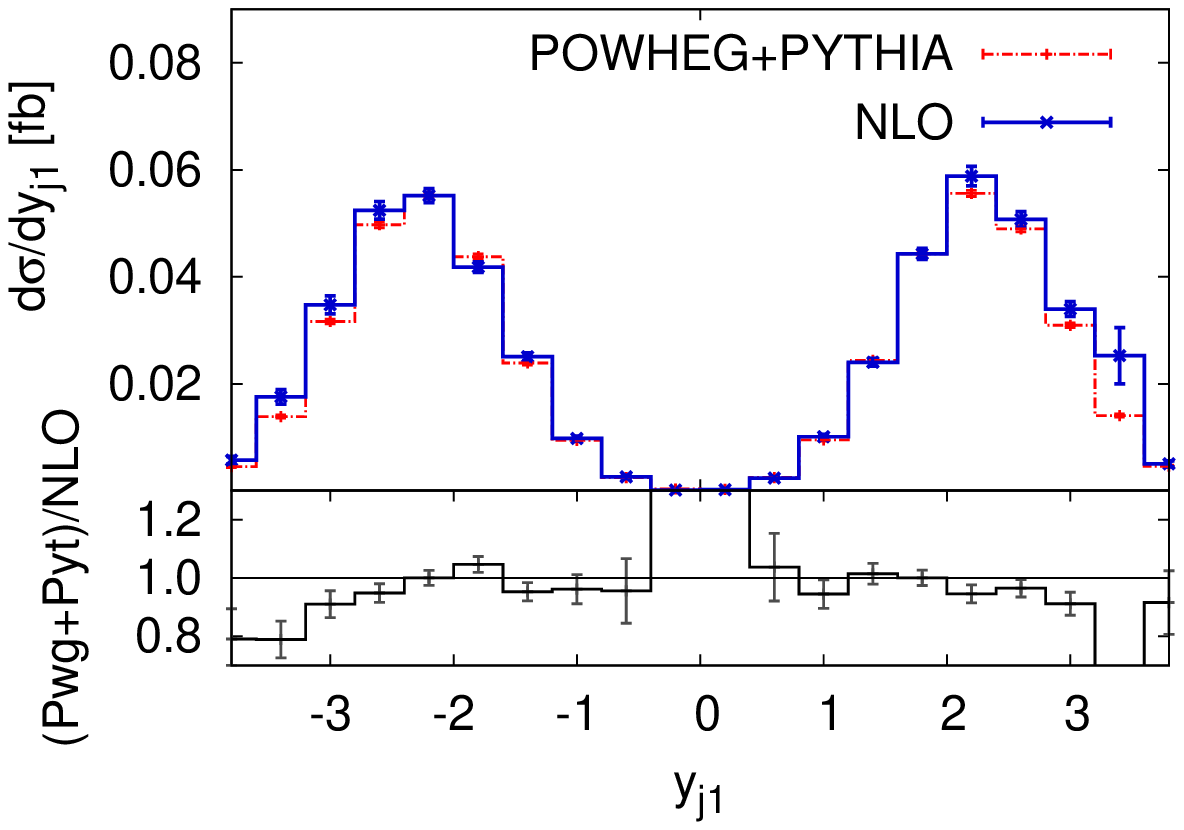}
\hspace*{1cm}
\includegraphics[angle=0,scale=0.5]{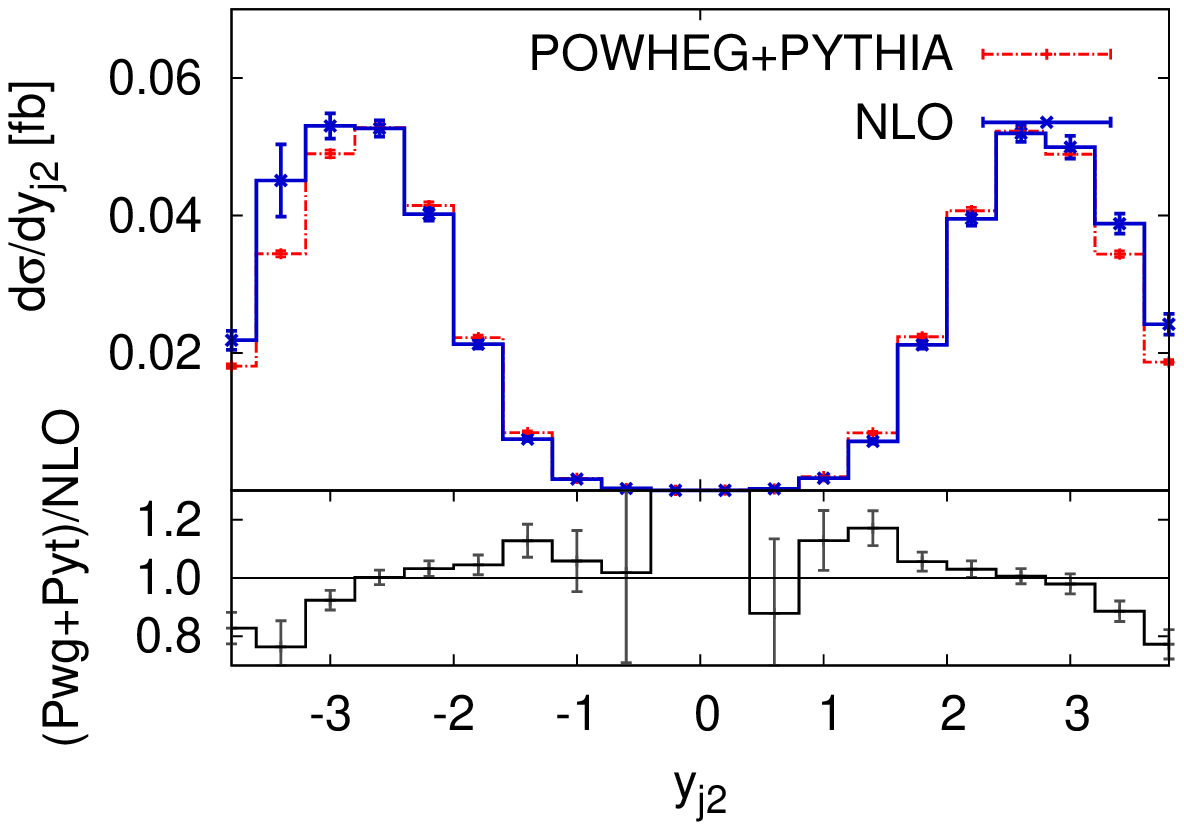}
\caption{Same as Fig.~\ref{fig:pttag} for the rapidity distribution of
  the hardest tagging jet (left) and second hardest tagging jet (right).}
\label{fig:ytag}
}

Good agreement between NLO and {\tt POWHEG+PYTHIA} is found for
observables related to the two hard charged leptons, such as their
transverse momentum and rapidity distributions, which are displayed in
Fig.~\ref{fig:ptl} for the positron. Since we neglect lepton-mass
effects in the NLO calculation, the corresponding results for the
$\mu^+$ are identical at NLO.
\FIGURE[t]{
\includegraphics[angle=0,scale=0.5]{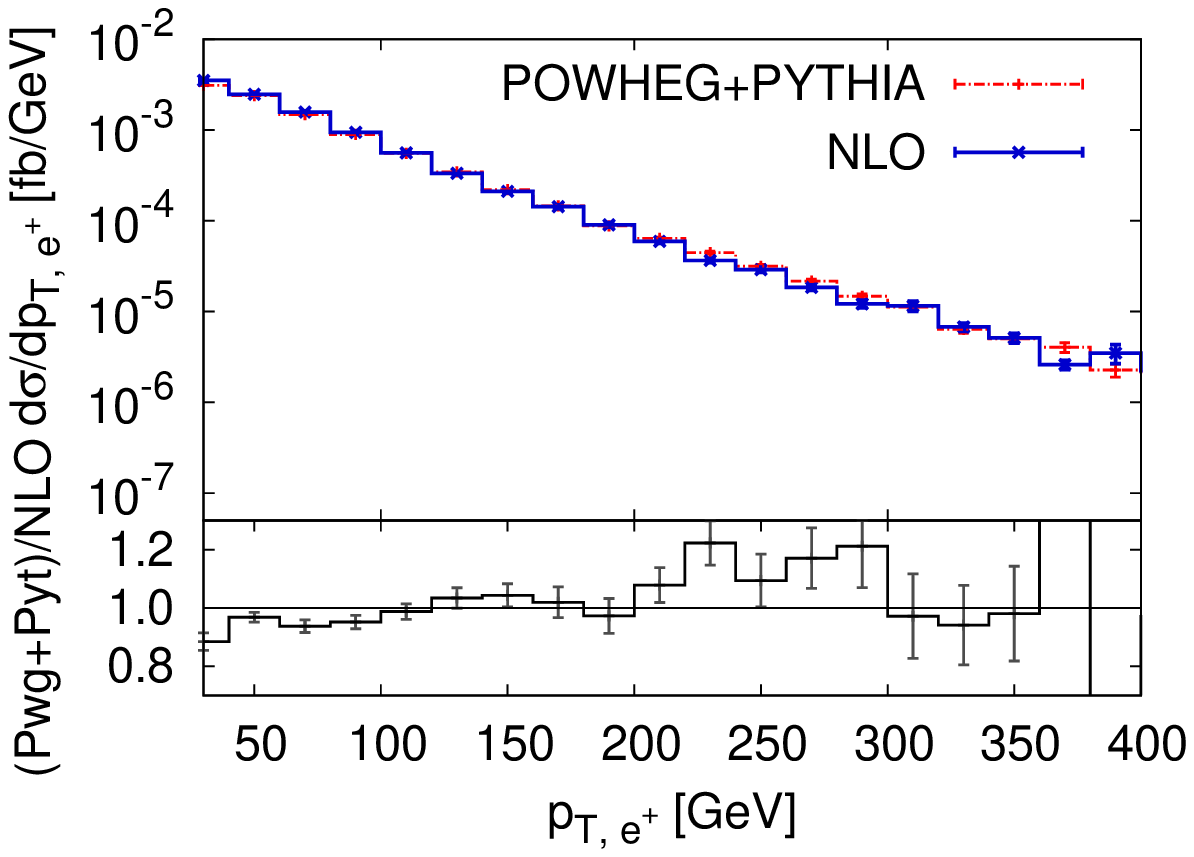}
\hspace*{1cm}
\includegraphics[angle=0,scale=0.5]{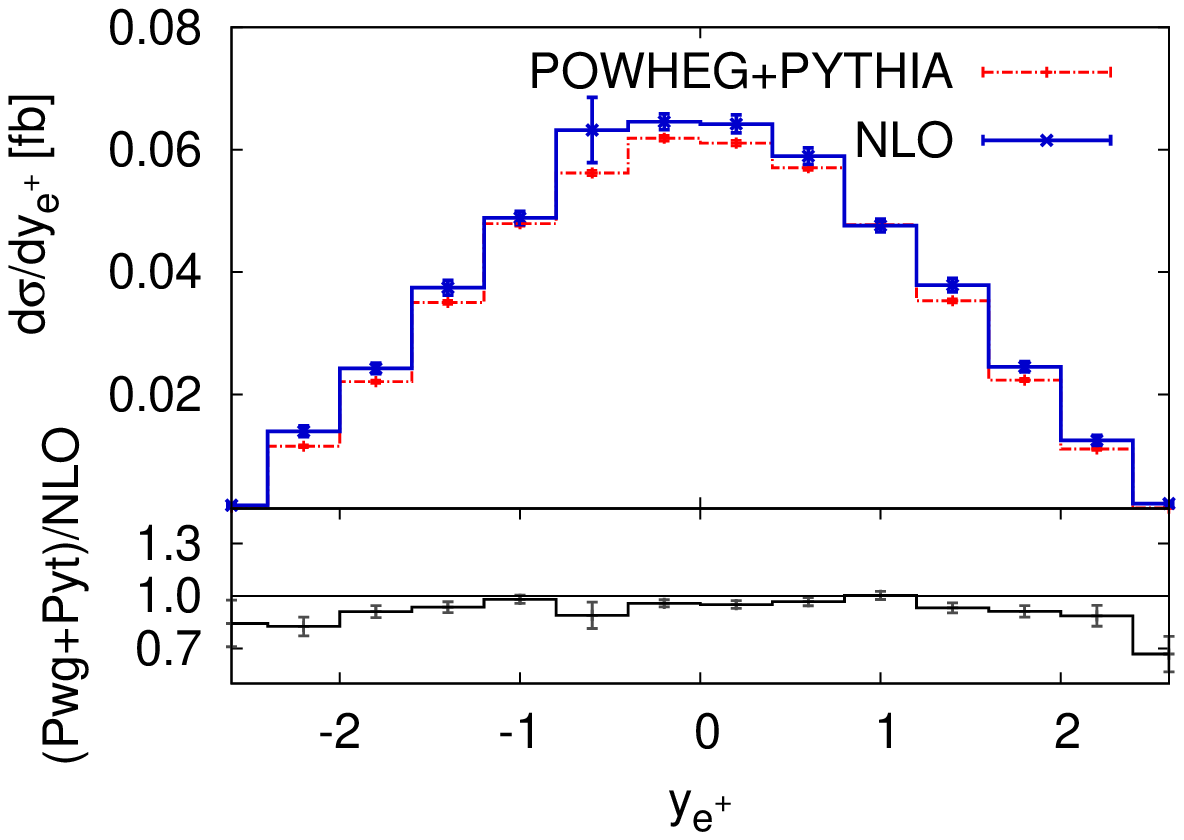}
\caption{Same as Fig.~\ref{fig:pttag} for the transverse momentum (left) and rapidity (right) distributions  of the positron.} 
\label{fig:ptl}
} 
While the shapes of these distributions barely change, their magnitudes 
are reduced in the same way as $\sigma^\mr{cuts}$ when the NLO
calculation is merged with {\tt PYTHIA}. This effect can also be
observed in the invariant mass distribution of the $e^+$ and the
$\mu^+$ and in the azimuthal angle separation of the two tagging jets,
shown in Fig.~\ref{fig:mll}.
\FIGURE[t]{
\includegraphics[angle=0,scale=0.5]{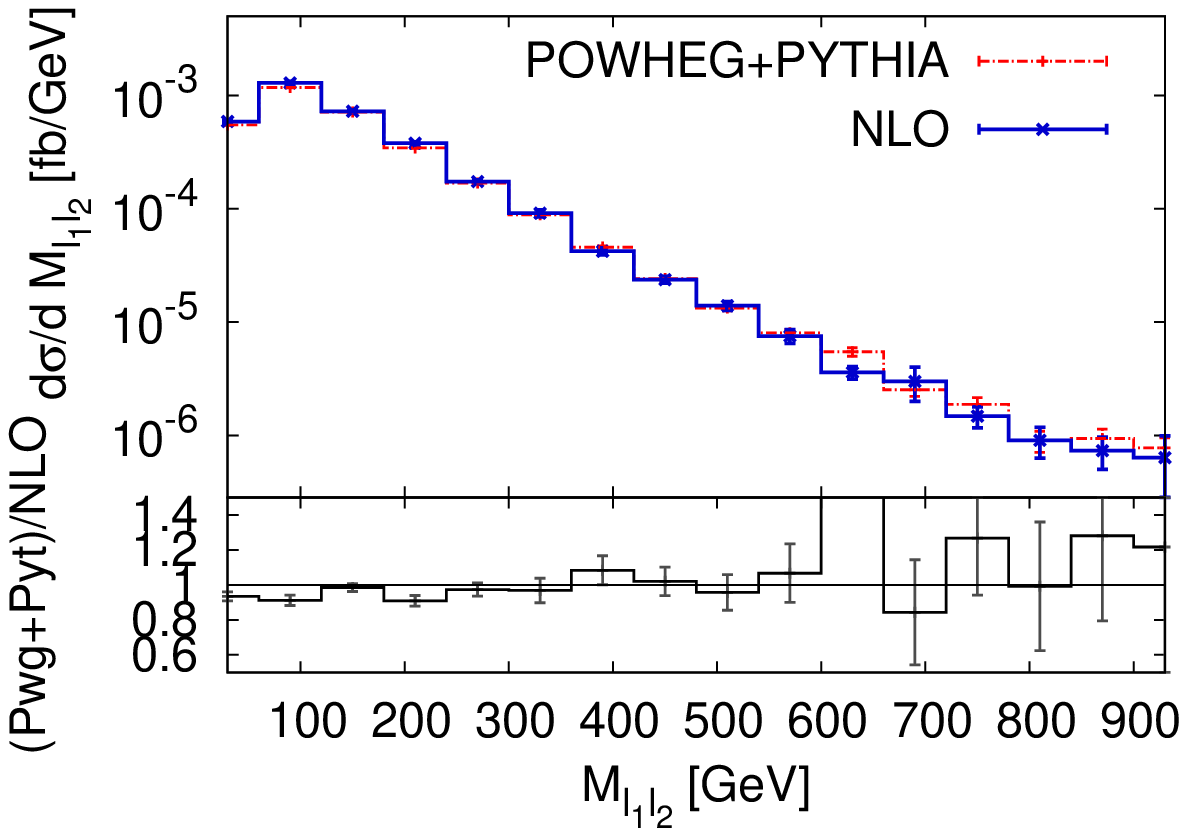}
\hspace*{1cm}
\includegraphics[angle=0,scale=0.5]{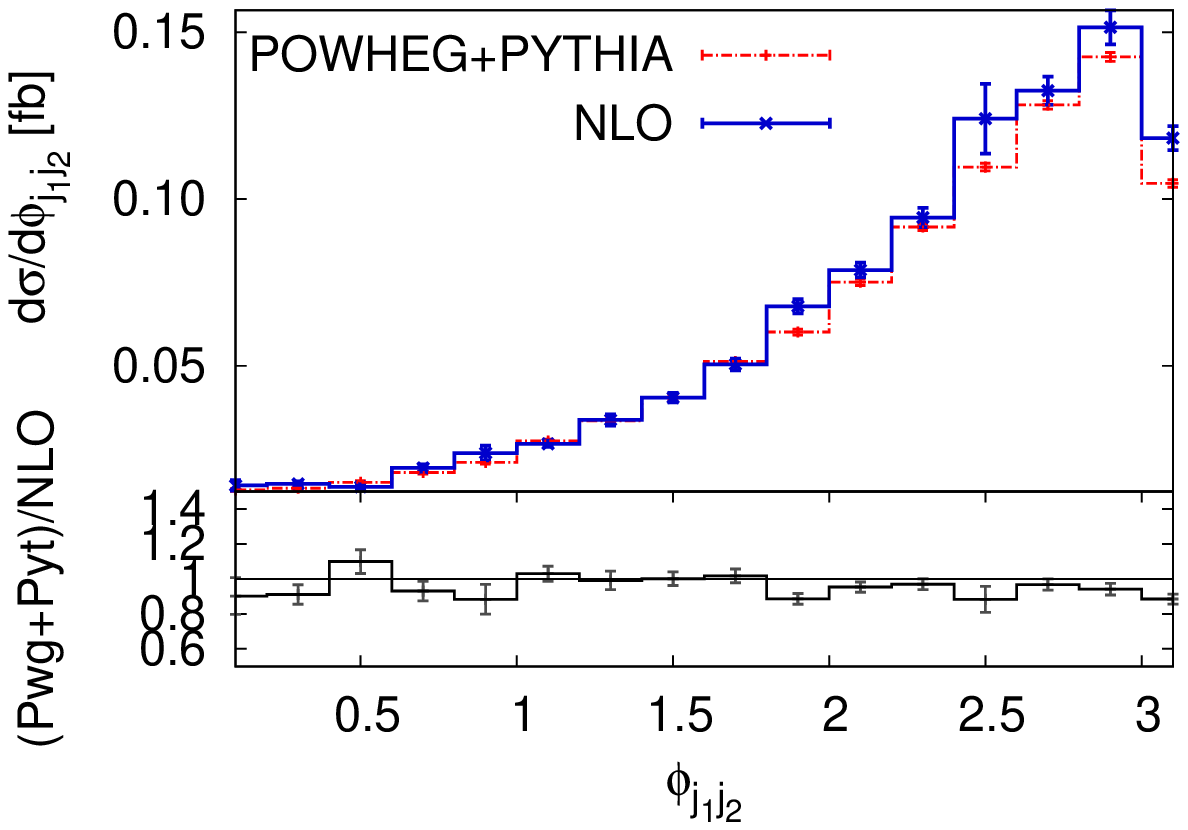}
\caption{Same as Fig.~\ref{fig:pttag} for the invariant mass
  distribution of the two charged leptons (left) and the azimuthal angle
  separation of the two tagging jets (right).}
\label{fig:mll}
} 

At NLO-QCD, a third jet can only be generated by real-radiation
contributions. All observables related to the third jet are therefore
effectively known at leading order only.  Furthermore, when the NLO
matrix element is matched with a parton-shower program, even if the
hardest emission is always generated by \POWHEG{}, a third jet can also
emerge from the parton shower. It is therefore interesting to compare
the fixed-order parton-level prediction with the \POWHEGpPYTHIA{} result for observables being particularly sensitive to this extra emission.
Figure~\ref{fig:pt3}~(left panel) shows that for the third jet transverse momentum distributions of very similar shape are obtained with the parton-level and the \POWHEGpPYTHIA{} programs. 
\FIGURE[t]{
\includegraphics[angle=0,scale=0.5]{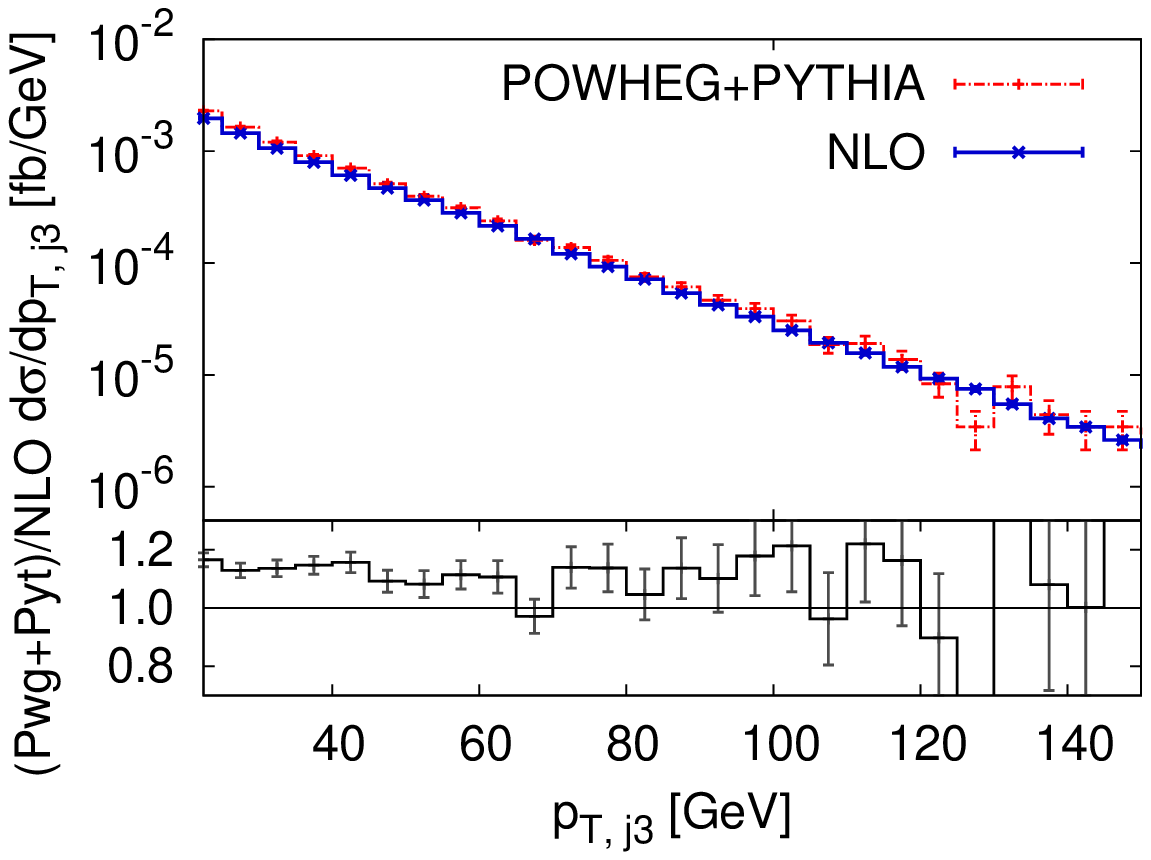}
\hspace*{1cm}
\includegraphics[angle=0,scale=0.5]{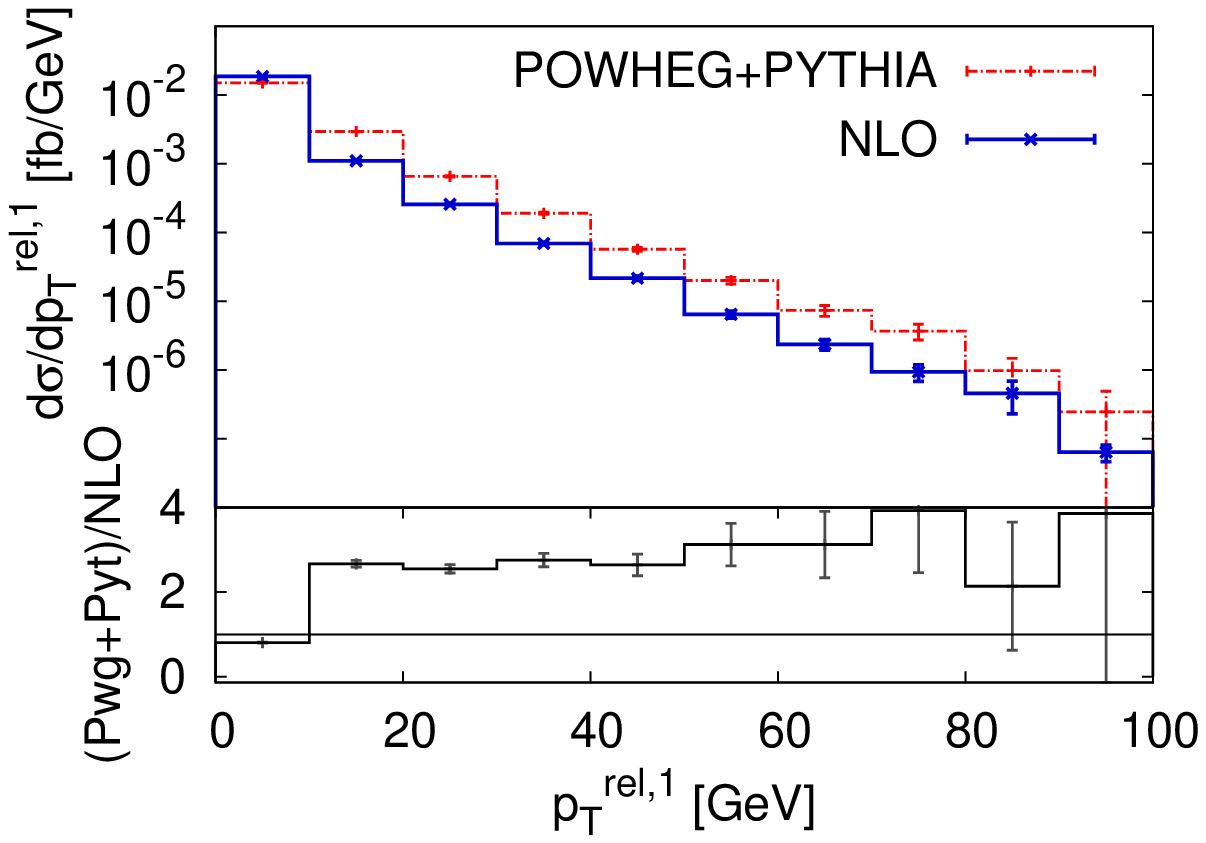} 
\caption{Same as
  Fig.~\ref{fig:pttag} for the transverse momentum distribution of the
  third jet (left) and relative transverse momentum distribution
  inside the hardest tagging jet (right).}
\label{fig:pt3}
} 

The relative distribution of the transverse momenta of all particles
inside a jet, defined with respect to the jet axis in the frame where
the jet has zero rapidity and momentum $p_j$, is given by
\beq
p_{T}^\mr{rel,j} = \sum_{i \in j}\frac{|\vec k_i \times \vec p_j|}{|\vec p_j|}\,. 
\label{eq:ptrel}
\eeq
Here, $k_i$ denotes the momentum of the $i^\mr{th}$ particle (track)
inside the jet.  Figure~\ref{fig:pt3}~(right panel) illustrates
$p_{T}^\mr{rel,j}$ for the hardest tagging jet.  At LO, only one
parton contributes to this jet. The jet axis is thus identical with
the associated parton track and $p_{T}^\mr{rel,1}$ vanishes. At NLO,
real-emission configurations give rise to non-zero values of
$p_{T}^\mr{rel,1}$. The largest contribution comes from partons which
are quasi-soft or collinear to the hard parton associated with the
jet, which results in a divergence of $d\sigma/dp_{T}^\mr{rel,1}$ at
$p_{T}^\mr{rel,1}=0$. The \POWHEGpPYTHIA{} prediction is damped in
this region, but lies above the pure NLO result everywhere else.

As already mentioned previously, contrary to QCD-mediated reactions,
in weak boson scattering processes jet activity in the
central-rapidity region is suppressed due to the color-singlet nature
of the weak boson exchange in the $t$-channel. This feature can be
exploited for background suppression by imposing a central jet veto,
i.e., discarding all events which exhibit substantial jet activity in
the rapidity region between the two tagging jets~\cite{Rainwater:1996ud}. 
For employing
central-jet veto techniques reliably in realistic simulations it is
crucial to quantitatively understand the impact of parton-shower
effects on jet emission in the central-rapidity range. In the context
of Higgs production via VBF, to this end detailed studies have been
performed~\cite{Figy:2003nv,Figy:2007kv,Hackstein:2007,Nason:2009ai}. 
Similarly to what happens in Higgs production via VBF, in $pp\to \wpp jj$ differences between the parton-level and the \POWHEGpPYTHIA{} results occur in the  rapidity distribution of the third jet. As illustrated by Fig.~\ref{fig:y3}~(left panel), 
\FIGURE[t]{
\includegraphics[angle=0,scale=0.5]{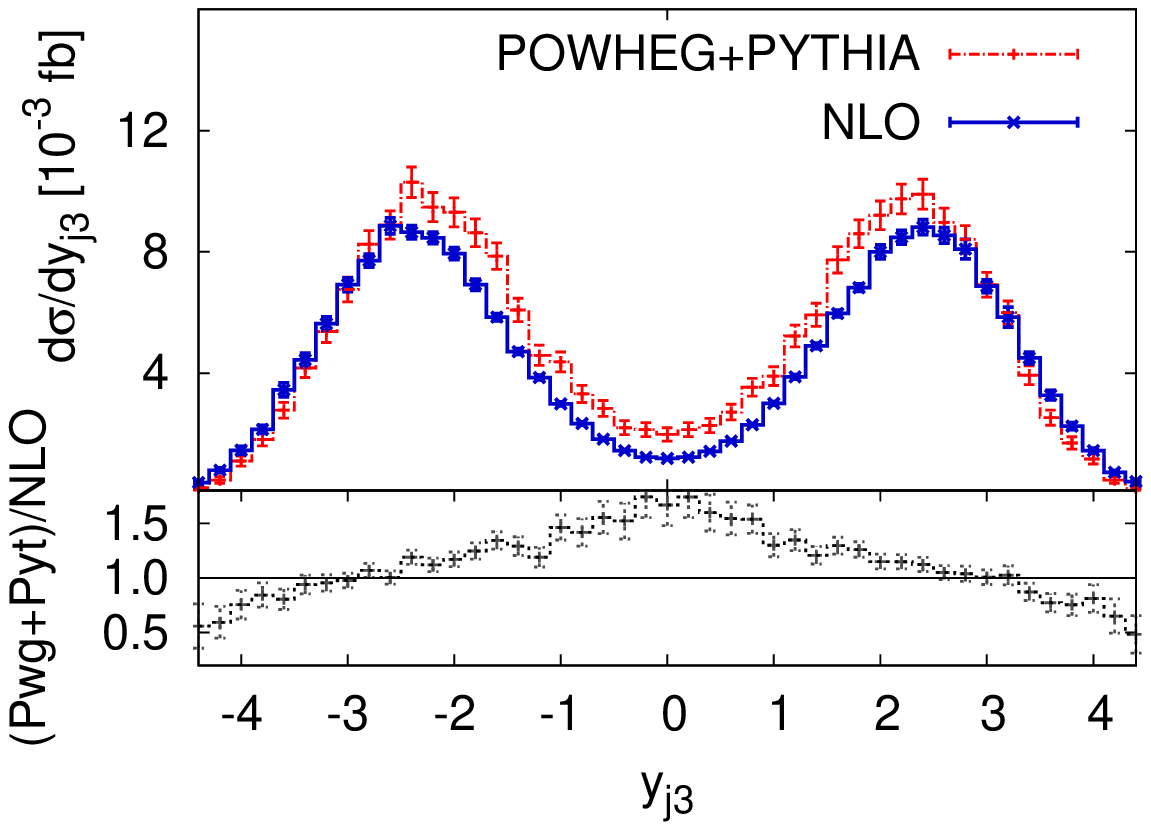}
\hspace*{1cm}
\includegraphics[angle=0,scale=0.5]{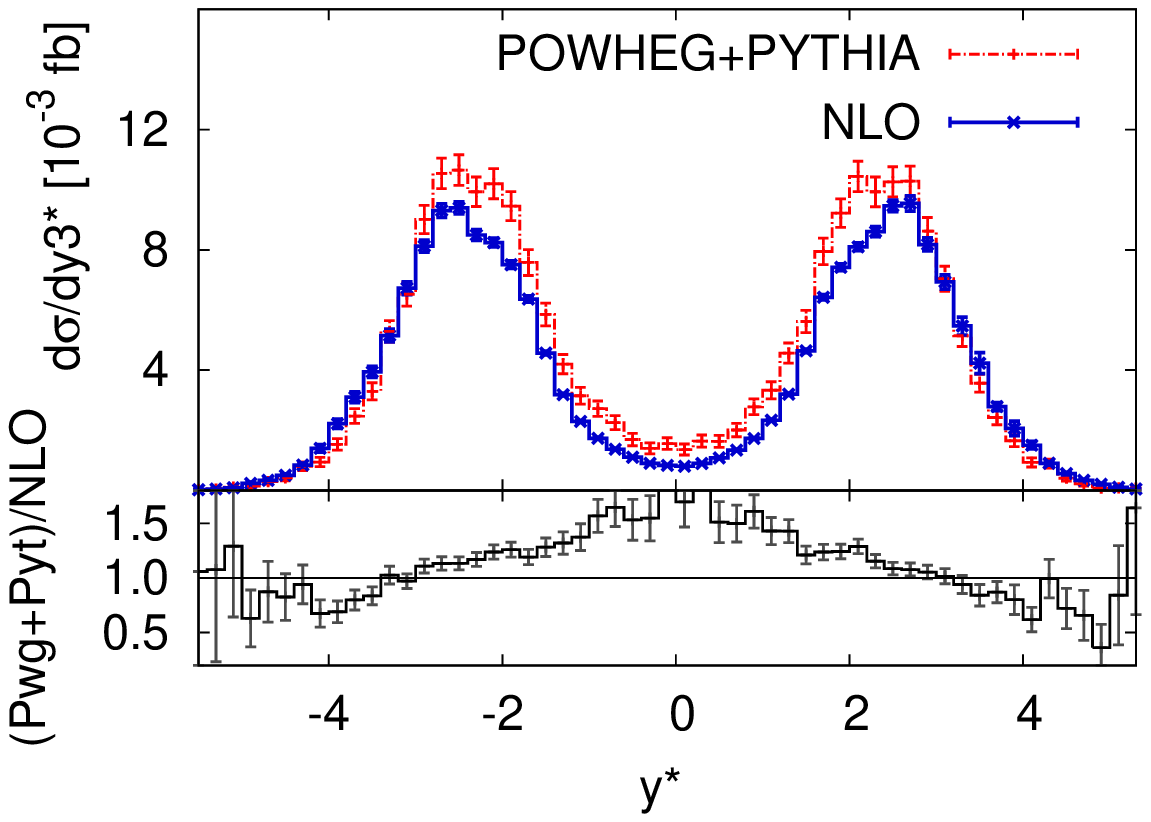}
\caption{Same as Fig.~\ref{fig:pttag} for the rapidity distribution of
  the third jet (left) and rapidity of the same jet with respect to
  the average of the rapidities of the two tagging jets, $y^\star$.}
\label{fig:y3}
}
in the pure NLO case a distinct dip occurs in $d\sigma/dy_{j_3}$,
which is partially washed out in the {\tt POWHEG+PYTHIA} result, thus
weakening the case of a central jet veto.

The position of the third jet with respect to the average of the two tagging jets can be accessed by the observable $y^\star$, which is defined as
\beq
y^\star = y_{j_3}-\frac{y_{j_1}+y_{j_2}}{2}
\eeq
and is displayed in Fig.~\ref{fig:y3}~(right panel).  When the third jet
is located in the central-rapidity region, $y^\star$ is close to zero,
while $|y^\star|$ values of about three indicate that the third jet is
close to one of the tagging jets, as is obvious from the position of
the peaks in the rapidity distributions of the two tagging jets (c.f.\
Fig.~\ref{fig:ytag}).  If a parton ends up close to
one of the tagging jets, it is likely to be recombined into this hard
jet rather than giving rise to an extra jet
\cite{Hackstein:2007}. From Fig.~\ref{fig:y3} it becomes apparent that
the extra hadronic activity due to the parton shower when \POWHEG{} is
matched with \PYTHIA{} is preferentially emitted in the region between
the two tagging jets. 

\section{Conclusion}
\label{sec:conc}
In this work, we presented the implementation of the NLO-QCD
calculation for electroweak $\wpp jj$ production in hadronic
collisions in the \POWHEGBOX~\cite{Alioli:2010xd}, a framework which allows
the merging of NLO-QCD calculations with parton-shower programs such
as {\tt PYTHIA} and {\tt HERWIG}. In this way the accurate
perturbative description of the hard partonic scattering event can be
combined in a well-defined manner with a realistic simulation of the
hadronic environment in $pp$ collisions.  Our implementation is
publicly available following the instructions provided at the \POWHEGBOX{}
web site {\tt http://powhegbox.mib.infn.it}.

With its distinct signature, the reaction $pp\to \wpp jj$ is
interesting not only from a technical point of view, but also as a
background for new physics searches and double parton scattering
processes. It receives contributions from QCD and EW production
modes. The QCD $\wpp jj$ production has already been considered in the
framework of the \POWHEGBOX{} in Ref.~\cite{Melia:2011gk}. Our new
implementation of the EW production mode in the \POWHEGBOX{} allowed
us to perform a phenomenological comparison of the two processes,
going beyond the estimates of Ref.~\cite{Melia:2010bm}. While naively
one could expect the QCD contributions to be much larger than the EW
ones, the QCD cross section is only approximately twice as large as
the EW one for inclusive cuts. VBF-specific selection cuts further
suppress the QCD contributions efficiently, resulting in a ratio
$\sigma^\mr{cuts}_\mr{QCD}/\sigma^\mr{cuts}_\mr{EW}\approx 0.04$. We
conclude that in the range where weak boson scattering is searched for
experimentally, no significant contamination from QCD-induced $\wpp
jj$ production is to be expected.

In this region, we have studied the impact of parton-shower effects on
NLO-QCD results for several characteristic distributions. We found
only a mild distortion of selected observables, such as rapidity
distributions of the hardest jets and leptons.  Larger differences
between the NLO parton-level and the \POWHEGpPS{} predictions can be
encountered, however, for the rapidity of the third jet. A
quantitative understanding of observables related to the third jet is
important if central-jet vetoing techniques are to be utilized at the
LHC.

{\bf Acknowledgments} 
We are grateful to Paolo Nason for helpful comments regarding the
implementation in the \POWHEGBOX{}. We thank Carlo Oleari and Dieter
Zeppenfeld for a careful proof-reading of the manuscript and for their
valuable feedback.  We furthermore would like to thank Gavin Salam and
Simon Pl\"atzer for useful discussions. The work of B.~J.\ is
supported by the Research Center {\em Elementary Forces and
  Mathematical Foundations (EMG)} of the
Johannes-Gutenberg-Universit\"at Mainz. G.~Z.\ is supported by the
British Science and Technology Facilities Council, by the LHCPhenoNet
network under the Grant Agreement PITN-GA-2010-264564 and by the
European Research and Training Network (RTN) grant Unification in the
LHC ERA under the Agreement PITN-GA-2009-237920.

%

\end{document}